# Optimum Signal Linear Detector in the Discrete Wavelet Transform - Domain

Ignacio Melgar Bautista✝
Jaime Gómez Saenz de Tejada*
Juan Seijas Martínez-Echevarría✝

Sener Ingeniería y Sistemas, S.A. ✝*
Departamento de Señales, Sistemas y Radiocomunicaciones, Universidad Politécnica de Madrid. ✝
Escuela Politécnica Superior, Universidad Autónoma de Madrid*
ignacio.melgar@sener.es
jaime.gomez@ii.uam.es
seijas@gc.ssr.upm.es

*Abstract:* - The problem of known signal detection in Additive White Gaussian Noise is considered. In this paper a new detection algorithm based on Discrete Wavelet Transform pre-processing and threshold comparison is introduced. Current approaches described in [7] use the maximum value obtained in the wavelet domain for decision. Here, we use all available information in the wavelet domain with excellent results. Detector performance is presented in Probability of detection curves for a fixed probability of false alarm.

*Key-words:* - Wavelet, signal detection, signal-to-noise ratio, probability of false alarm, probability of detection.

## 1 Introduction

Wavelet transform is one of the most successful methods in signal de-noising processes.

Let **s** be a sampled signal with energy contained in a certain range of frequencies, and let **n** be sampled noise alone. Using Subband Coding Algorithm [3], its wavelet coefficients can be obtained for different levels of decomposition. The wavelet coefficients with greater signal to noise ratio (SNR) will be those which correspond to frequency scales where signal energy is concentrated. In [8] a signal detection algorithm is proposed based on comparing the component with maximum absolute value of the Discrete Wavelet Transform (DWT) coefficients for a given scale with a certain threshold. Given two sample sets, N = {**n**} and S = {**s**}, and let c be the DWT scale, then

$$\mathbf{d}_n = DWT_c(\mathbf{n}), \quad (1)$$
$$u_n = \text{Max-Abs}(\mathbf{d}_n). \quad (2)$$

Considering a fixed probability of false alarm ($P_{fa}$), a threshold $V_T$ is calculated obtained using Montecarlo estimation for set N and condition $V_T \leq u_n$. Then probability of detection ($P_D$) is computed for each SNR, showing the best scale overall,

$$\mathbf{d}_s = DWT_c(\mathbf{s}), \quad (3)$$
$$u_s = \text{Max-Abs}(\mathbf{d}_s), \quad (4)$$

and using Montecarlo for set S and condition $V_T \leq u_s$.

Note that in this algorithm, only one component in the wavelet domain is being used. This component is not necessarily set in the process, and it is certainly the most relevant. But it is still only one. A big amount of information is being rejected, that of the other not-so-important components in vector **v**.

In figure 1, we present an example where the mean of every component in the vector alone is compared for both N and S sets. There are some peaks that are the best candidates for being over the threshold, but as the statistics deviation gets higher than the mean (worst SNR), it gets much more difficult to find one component clearly over maximum noise. Even though there could be many components close to the threshold, which is very unlike noise DWT figures, this information is not used. Moreover, negative peaks are treated as positive using the absolute value function. This process loses information about each component.

To distinguish between this algorithm and our own, the one described in [8] will be referred as *Max-Detector* throughout this paper, and the new one as *Linear-Detector*.

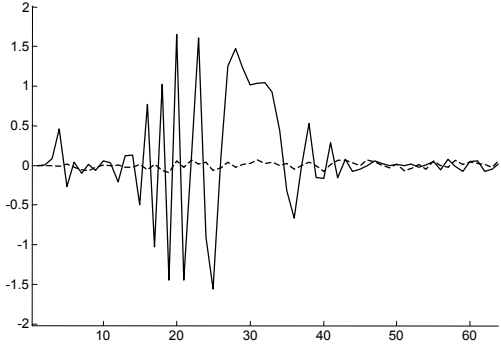

Fig.1: Detailed mean wavelet coefficients for noise (dashed line) and signal (solid line), using the wavelet and chirp parameters identified in section 5 of this paper. SNR = –5 dB, scale = 16. Mean is computed averaging 500 noise experiments.

## 2 Wavelet transform

Almost all existing signals can be expressed by a wavelet transform. Wavelets are generated by the scale and translation of a single prototype function called wavelet mother:

$$\psi_{s,\tau}(t) = \frac{1}{\sqrt{|s|}} \psi\left(\frac{t-\tau}{s}\right), \qquad (5)$$

where $\psi$ is the mother wavelet, s is the scaling factor and $\tau$ is the translation factor. They are building blocks of wavelet transform for different scales and translations, just as trigonometric functions of different frequencies are building blocks of Fourier transform [6].

In the case of discrete wavelet transform (DWT), $\tau$ and s also take discrete values, given by

$$s = s_0^m, \qquad (6)$$

$$\tau = n\tau_0 s_0^m, \qquad (7)$$

$$\psi_{m,n}(t) = s_0^{-m/2} \psi\left(s_0^{-m} t - n\tau_0\right), \qquad (8)$$

A particular class of wavelets are orthonormal wavelets which are linearly independent, complete and orthogonal. This means that there is no "redundant" data from the original signal in more than one wavelet. In [2] Daubechies developed conditions under which wavelets form orthonormal bases. Thus the Discrete Wavelet Coefficients are the inner products of the signal and wavelet function. That is:

$$f(t) = \sum_{m,n} \psi_{m,n}(t) <\psi_{m,n}(t), f(t)>, \qquad (9)$$

Multiresolution analysis allows to look a signal at different scale by "zooming in " or "zooming out"; that is, an approximation of a given signal at low resolution can go to an approximation at immediate higher resolution just by adding some "details" information.

In [3] Mallat developed a fast wavelet algorithm based on the pyramid algorithm of Burt and Adelson [1]. The basic components in each stage of the pyramid are two analysis filters: a low-pass filter **h** and high-pass filter **g**, and a decimation by two operation. As it can be easily observed, because of the Heisenberg uncertainty principle; DWT offers high time resolution for low scales (high frequencies) and high frequency resolution for high scales (low frequencies).

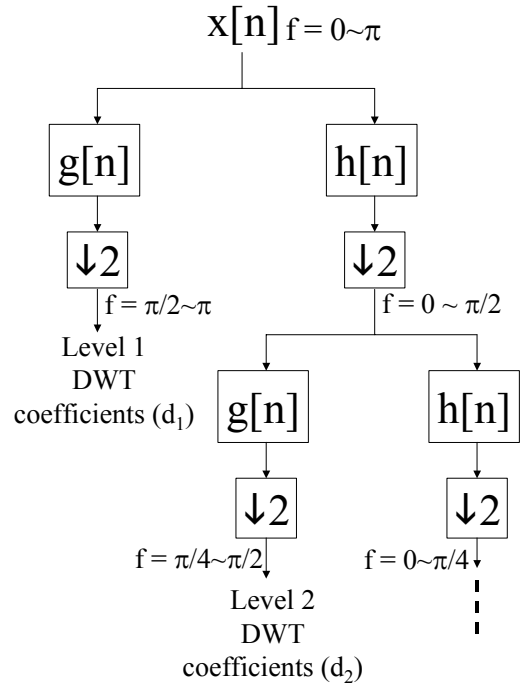

Fig.2 The dyadic decomposition algorithm

## 3 Linear-Detector

### 3.1 Algorithm description

In the algorithm defined with equations (1) to (4), we must change the inefficient way information is integrated into one only real number. The steps defined with equations (2) and (4) now become:

$$u_n = \sum_{i=1}^{M} a_i d_{ni}, \qquad (10)$$

$$u_s = \sum_{i=1}^{M} a_i d_{si}, \quad (11)$$

being M the number of components in vector **v**, and $a_i \in \Re$ the coefficients (or weights) for each component. Montecarlo estimation of $P_{fa}$ and $P_D$ is no longer needed if we use the algorithm described in section 3.3. These coefficients can be calculated with the optimum-finding algorithm. By defining the Linear-Detector we are now having n times more components with information to classify pulses form noise, increasing the $P_d$ for all values of SNR. Note that it can be shown that the Linear-Detector approach is, at least, as good as the Max-Detector approach, and it usually behaves much better.

### 3.2 Theoretical comparison

The objective of this theoretical comparison is to show that we can always find a set of coefficients $a_i$ such that the performance of our method is equal or better than the former approach. We will use subindex M and L to denote Max-Detector and Linear- Detector methods.

For the sake of clarity let us suppose we have a vector **v** with two components only. Initially our algorithm sets no constraint to the coefficients, but for the ease of comparison we will introduce one:

$$\sum_{i=1}^{M} a_i^2 = 1, \; i \in [1,M]. \quad (12)$$

We will also simplify this development by allowing positive peaks only, i.e., $\mu_i \geq 0$ and $\sigma_i=1$. Suppose we set the threshold $V_T$ to some value rendering $P_{fa}$ small, let's call $P_{faX}$ to the $P_{fa}$ for one component alone, then

$$P_{faM} = 2\, P_{faL} = 2\, P_{faX}. \quad (13)$$

Information can be distributed anyhow between the two components, and the efficiency for each method is defined as monotonic while the quantity of information remains unchanged, i.e., $\mu_1+\mu_2$ is constant. The mean for our distribution is

$$\mu_L = a_1\, \mu_1 + a_2\, \mu_2. \quad (14)$$

It can be easily shown that the optimum value for $\mu_L$ occurs when the following condition holds:

$$\frac{a_1}{a_2} = \frac{\mu_1}{\mu_2} \quad (15)$$

Now we have to analyse two cases:

- Only one component has information, for instance $\mu_1$; then $\mu_2 = 0$, $a_2 = 0$ and $a_1 = 1$. Therefore both methods are equivalent, even though $P_{dM} > P_{dL}$ using the previously fixed $V_T$.
- Both components have the same quantity of information. Then $\mu_1 = \mu_2$, $a_1 = a_2 = 2^{-1/2}$, and $\mu_L = 2^{1/2}\mu_1$. If there were no information at all, i.e., $\mu_1 = \mu_2 = 0$, then the previous case applies. But as we increase the information in the vector, it can be shown that the difference on probability of detection $P_{dM} - P_{dL}$ decreases (it can be negative) in a monotonic function until the limit is reached ($P_{dM}$ or $P_{dL}$ is 1).

Note that $P_{fa}$ does not change during all this development. From an equivalence point ($\mu_1 = \mu_2 = 0$), as the information increases in the vector (it cannot be decreased any more), $P_{dL}$ improves faster than $P_{dM}$. On the other hand, as we go from the first case to the second case, the Linear-Detector approach worsens slower than the Max-Detector. Therefore, the initial assumption holds for the 1- and 2-component cases.

Intuitively, if the n-1-component case holds, the n-component case also holds. Note that the n-component case can be divided in different subproblems of size n-1, n-2, etc.. down to 1, plus one subproblem having all n components with the same quantity of information. This subproblem can be solved using a similar approach, with an initial difference $P_{dM} - P_{dL}$ being smaller (including negative numbers) as n increases.

### 3.3 Optimum-finding algorithm

In this subsection we describe the algorithm used to determine the presence of a known signal inside additive white Gaussian noise (AWGN).

Using vector notation for signal representation, let **v** be defined as a $2^N$ length vector containing a sampled pulse in the presence of AWGN. Noise distribution is zero-mean and $\sigma_n^2$ variance. For the true hypothesis ($H_1$) vector **v** is the addition of sampled noise and sampled signal with a certain signal-to-noise ratio (SNR). For the false hypothesis ($H_0$) signal is not present, thus **v** contains only noise. Therefore **v** can be expressed, using natural scale, as follows:

$$\mathbf{v}|_{H_0} = \mathbf{n}, \quad (16)$$

$$\mathbf{v}|_{H_1} = 10^{\frac{SNR}{20}} \sigma_n^2 \hat{\mathbf{s}} + \mathbf{n}, \quad (17)$$

where $\hat{\mathbf{s}}$ is the sampled pulse sequence with power normalized to 1, $\mathbf{n}$ is the sampled AWGN and SNR is expressed in dB units.

Using Subband Coding algorithm [3] we obtain $\mathbf{d}_{i_{(s)}}$ vector as the i-th level detail wavelet coefficients of $\hat{\mathbf{s}}$. $\mathbf{d}_{i_{(s)}}$ vector length M is given by

$$M = 2^{N-i}. \qquad (18)$$

Assuming that high-pass and low-pass filter impulse responses h[n] and g[n] are normalized to unitary energy, that is

$$\sum_{n=0}^{L-1}(h[n])^2 = \sum_{n=0}^{L-1}(g[n])^2 = 1, \qquad (19)$$

being L the filter length, then i-th wavelet coefficients of vector $\mathbf{v}$ can be written as

$$\mathbf{d}_i\big|_{H_0} = \mathbf{d}_{i_{(n)}}, \qquad (20)$$

$$\mathbf{d}_i\big|_{H_1} = 10^{\frac{SNR}{20}}\sigma_n^2 \mathbf{d}_{i_{(s)}} + \mathbf{d}_{i_{(n)}}. \qquad (21)$$

$\mathbf{d}_{i_{(n)}}$ vector corresponds to i-th detail wavelet coefficients of $\mathbf{n}$ and length given by (18). Each k-th component of $\mathbf{d}_{i_{(n)}}$ is a zero-mean and $\sigma_n^2$ variance gaussian random variable for $\lceil L/2 \rceil \le k \le M$ (when the filter reaches the steady stage).

Let us introduce the variable $\upsilon$. Pulse presence will be asserted if the value of $\upsilon$ is greater than a certain threshold ($V_T$). Now let define $P_{fa}$ and $P_D$ as

$$P_{fa} = \Pr\{y > V_T/H_0\}, \qquad (22)$$

$$P_D = \Pr\{y > V_T/H_1\}. \qquad (23)$$

For a given vector $\mathbf{a} \in \Re^M$, $\upsilon$ is created by

$$\upsilon = \sum_{k=\lceil L/2 \rceil}^{M} \mathbf{d}_i[k]\mathbf{a}[k]. \qquad (24)$$

So, $\upsilon$ is a gaussian random variable with $N(\eta_\upsilon, \sigma_\upsilon)$ normal distribution, where

$$\eta_\upsilon\big|_{H_0} = E\{y\}\big|_{H_0} = \sum_{k=\lceil L/2 \rceil}^{M} \mathbf{a}[k] E\{\mathbf{d}_{i_{(n)}}[k]\} = 0, \qquad (25)$$

$$\eta_\upsilon\big|_{H_1} = E\{y\}\big|_{H_1} = \sum_{k=\lceil L/2 \rceil}^{M} \mathbf{a}[k] E\Big\{10^{\frac{SNR}{20}}\sigma_n^2 \mathbf{d}_{i_{(s)}}[k]$$

$$+ \mathbf{d}_{i_{(n)}}[k]\Big\} = 10^{\frac{SNR}{20}}\sigma_n^2 \sum_{k=\lceil L/2 \rceil}^{M} \mathbf{a}[k] E\{\mathbf{d}_{i_{(s)}}[k]\} \qquad (26)$$

and

$$\sigma_\upsilon^2\big|_{H_0} = \sigma_\upsilon^2\big|_{H_1} = \sigma_n^2 \sum_{k=\lceil L/2 \rceil}^{M}(\mathbf{a}[k])^2 +$$

$$2 \sum_{k=\lceil L/2 \rceil}^{M-1}\left[\sum_{j=k+1}^{M} E\{\mathbf{a}[k]\mathbf{a}[j]\mathbf{d}_{i_{(n)}}[k]\mathbf{d}_{i_{(n)}}[j]\}\right]. \qquad (27)$$

The second term in the sum of (27) is negligible with respect to the first term. Therefore we simplify that expression to the following one:

$$\sigma_\upsilon^2\big|_{H_0} = \sigma_\upsilon^2\big|_{H_1} = \sigma_n^2 \sum_{k=\lceil L/2 \rceil}^{M}(\mathbf{a}[k])^2. \qquad (28)$$

As it is shown in (25), (26) and (28), for a given $P_{fa}$ and vector $\mathbf{a}$ with at least one non-zero component, the associated $V_T$ can be obtained. If SNR is also considered then $P_D$ can be computed as well. Therefore it is possible to reach maximum detection performance of the scheme proposed in (24) only by maximizing $P_D$ as a function of SNR and $\mathbf{a}$, for the previously fixed $P_{fa}$.

The proposed algorithm can be extended to improve probability of detection. Let us suppose detail wavelet coefficients of a sampled pulse for a set of K scales $B = \{i_k \mid 1 \le i_k \le N\}$ are computed. Then all of them are concatenated on a single vector $\mathbf{d}_B$, created as follows:

$$\mathbf{d}_B = [\mathbf{d}_{i_1}, \mathbf{d}_{i_2}, ... \mathbf{d}_{i_K}]. \qquad (29)$$

Now let us use the detection process proposed in (24) with $\mathbf{d}_B$. For a given $\mathbf{a}$ vector with same length as $\mathbf{d}_B$, (26) and (28) show $\sigma_\upsilon^2$ (and consequently $P_{fa}$) do not depend on set B, but $\eta_\upsilon$ and $P_D$ do. Therefore, if we only include on set B the wavelet coefficients for the scales where the sampled pulse has high amplitude in any of its components, then $\eta_\upsilon$ and $P_D$ will be increased. That means the chosen $i_k$–th scale wavelet coefficients will correspond to the discrete frequency ranges $[\pi/2^{i_k}, \pi/2^{i_k-1}]$ where the pulse has greater energy.

### 3.4 Algorithm complexity
The computational load for this algorithm depends on the phase.
For the optimum-finding coefficients calculation, resources are mostly used in whatsoever constrained-function-maximization algorithm is used. Nevertheless, this process is done only once, in the engineering process, and its performance is not critical.

For the operative phase, the new algorithm has to calculate the wavelet transform first, and then the linear function. Being N the initial vector size, and M the wavelet filter size, the number of multiply-add operations performed in the wavelet transform is 2MN. The number of multiply-add operations needed in the linear function calculation is N. Therefore, the computational resources needed for the complete algorithm are basically the same as for the wavelet transformation calculation, so the complexity remains O(MN).

## 4 Experimental results

The data used to confirm the theoretical development had the following features: chirp pulse, 1024 samples (see Fig.3); mother wavelet Daubechies 9, using $d_3$, $d_4$, $d_5$ and $d_6$ wavelet coefficients. For all experiments we used white Gaussian noise with zero mean and deviation equals one, and $P_{fa} = 10^{-3}$.

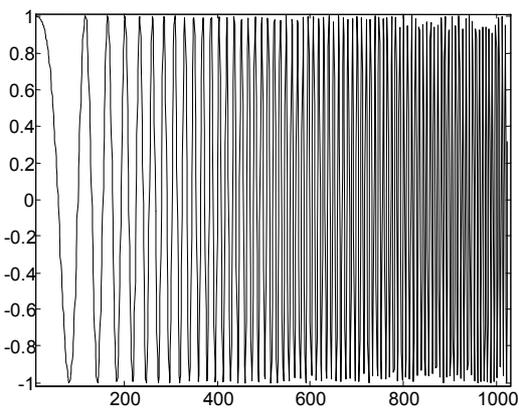

Fig.3: Chirp pulse with 1024 samples.

It can be observed in Fig.4 that Linear-Detector approach performs much better than the Max-Detector one. This effect increases notably with the filtered vector size. Information density as a whole remains unchanged, but as the vector size decreases, the resolution cell of each component increases. Therefore distribution gets less uniform and the probability for one component to draw together enough information increases.

In a previous section it was established that the Linear-Detector is always better or equal to the Max-Detector using the same information. But in fact, the Linear-Detector uses more information from the same source: the sign of negative peaks remains. The advantage is clear, although we have not yet calculated the performance upgrading.

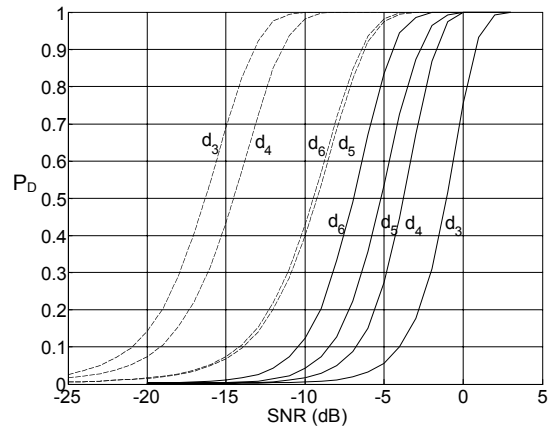

Fig.4: Comparison of different wavelet scales using the Max_Detector algorithm (solid lines) and Linear-Detector theoretical limits (dotted lines). $P_{fa}$ is always set to $10^{-3}$.

These two differences (information integration and negative sign use), are the main advantage for the new method. But there is also another important side effect. Using the Max-Detector method, you must guarantee all components have about the same normalized distribution, so that you can use a single useful threshold. This constraint cannot be assured when you want to use together more than one wavelet scale of the same pulse. Using the Linear-Detector approach, you can use together more than one scale, more than one wavelet transform, or even more than one kind of filter. Components in the vector may have different orders of magnitude, but they will all be treated as equally-sized using the coefficients. The more non-redundant information you use, the better probability of detection you will get. There is no theoretical limit to the quantity of information you can draw together.

In Fig.5 the results for a simple multiple-source representation are shown. Using the same basic parameters as in the previous experiments we concatenated the results of applying $d_3$, $d_4$, $d_5$ and $d_6$ wavelet coefficients in one single vector. The probability of detection increases dramatically compared to the best single scale.

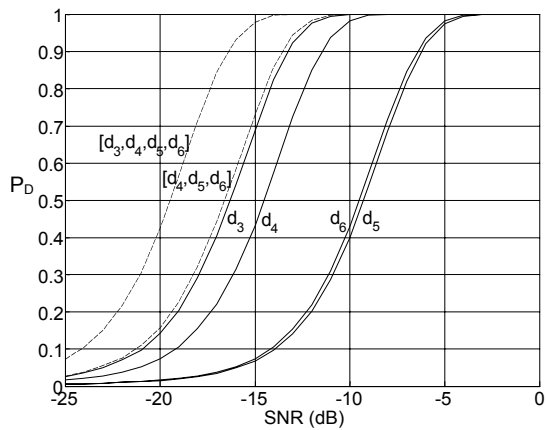

Fig.5: Comparison of two multi-scale vector, (leftmost is concatenation of $d_3$, $d_4$, $d_5$, and $d_6$, and the following is concatenation of $d_4$, $d_5$, and $d_6$), using the Linear-Detector, against the single-scale Linear-Detector theoretical limits. $P_{fa}$ is always set to $10^{-3}$.

## 5  Conclusions

A new algorithm has been introduced to integrate information for pulse detection much more efficiently than current approaches. In this paper we have presented some mayor differences: we observed that the Linear-Detector algorithm is always equal or better than the Max-Detector using the same information, and it outperforms as the vector size increases with energy-filled components; because of the component sign treatment, in normal circumstances the new approach uses more information than the former method; using the Linear-Detector algorithm, you have the possibility to integrate multiple-source information, with excellent results.

The results given here for the new approach correspond to theoretical reachable optimum coefficients. Nevertheless, these values cannot be implemented because they depend on SNR, which is unknown in the signal processor. Another algorithm is needed to find one only set of coefficients that is optimum for all SNR scenarios. We already have some preliminary results using a machine learning method: Support Vector Machines. Our first results are very close to the theoretical optimum, and they will be described in a soon-to-publish paper.

As it was established in this paper, there is no limit to the quantity of information that can be used to detect a pulse. An interesting line of research is open to find different combinations of filters giving better results. Note that, even though we used a specific wavelet transform and chirp pulse, there is no constraint in the algorithm about these parameters.

## 6  Acknowledgements

The authors would like to thank Joaquin Torres and Carlos Miravet for helpful discussions and paper review. This project is funded by Sener Ingeniería y Sistemas, in the frame of the Aerospace Division R&D program.